# A Procedure for Extracting Software Development Process Patterns

Mahdi Fahmideh, Fereidoon Shams
Automated Software Engineering Research Group, Electrical and Computer Engineering Faculty, Shaid Beheshti University GC, Tehran, Iran
{m_fahmideh, f_shams}@sbu.ac.ir

*Abstract*—Process patterns represent well-structured and successful recurring activities of Software Development Methodologies (SDMs). They are able to form a library of reusable building blocks that can be utilized in Situational Method Engineering (SME) for constructing a custom SDM or enhancing an existing one to fit specific project situation. Recently, some researchers have subjectively extracted process patterns from existing SDMs based on cumulative experience in various domains; however, how to objectively extract process patterns from SDMs by adopting a systematic procedure has remained as question. In this regard, this paper is concerned with a procedure aiming to take process patterns out of existing SDMs. An example illustrates applicability of the proposed procedure for extracting process patterns in a specific context.

*Keywords- Process Patterns, Situtional Method Engineering, Assembly-Based Method Engineering, Software Development Methodologies, Pattern Mining.*

## I. INTRODUCTION

A pattern is " a general solution to a common problem or issue, one from which a specific solution will be derived" [1, 2]. In software engineering, many types of patterns have already well-known, for instance GoF Design Patterns [3] and Gov Architectural Patterns (GoV) [4]. *Process Pattern* is a kind of pattern by which classes of common successful practices and recurring activities in specific SDMs are represented [2]. Typically, a SDM is consisted of two main parts [5]: a *Process* that is contained a set of activities, techniques, guidelines, principles, artifacts, roles, and tools for effective software development; and a *Modeling Language* to represent produced artifacts. Process patterns are the results of applying abstraction to recurring activities to form an effective mechanism for highlighting ones that have proven to be successful in SDMs. The process patterns are intended to be reused in SDMs. They enable method engineer to describe and document domain specific knowledge in SDMs in an abstract, well-defined, and maintainable structure. The main application of process patterns is in *Situational Method Engineering* (SME) specially *Assembly-Based Method Engineering* [6] approach in which process patterns form a rich library of reusable building blocks as called method chunks for constructing a custom SDM or enhancing an existing one to fit specific project situation at hand. A method chunk is viewed as an autonomous and coherent part of a SDM [7]. For instance the activities such as *Requirements Elicitation*, *Use-case Modeling* or *Develop Architecture* are considered as method chunk.

The process patterns open the areas of formal and quantitative measurement of software process that leads to applying analytical processes in SDMs [8]. Process patterns provide well-structured software process for organization's projects in general. Moreover, it represents the common conceptual base of a company's SDM to improve and evolve their development process [8].

There are various types of SDMs for developing software systems in different domains. For instance, in the domain of object-oriented system development OPEN, Booch, Objectory, OOSE, BON, Catalysis, USDP and RUP are the famous ones. Furthermore, SCRUM, DSDM, Crystal Clear, dX, FDD, and XP have emerged to support software development [9]. Each SDM prescribes its successive activities for developing target system. While all of the SDMs belong to a specific domain have same philosophy and concepts in software development, hence recurring activities might be repeated by different names. For instance, all the agile SDMs emphasize on three imperative activities generically called *Product Review*, *Process/Plan Review* and *Post-mortem Review* [10]. Similarly, most of the component-based SDMs have the same activities such as *Component Identification* and *Component Adaptation* [11].

The importance of process patterns will become more significant when method engineer faces excessive number of different SDMs. Since none of the SDMs can cover all the relevant issues in software development, it is difficult for method engineer to select appropriate practices to fit the project requirements. In addition, most of the SDMs in specific domain prescribe different activities with various names also they represent same activities yet from a different viewpoint. Therefore, selection of an appropriate SDM will be a serious issue in this situation. Process Patterns come up with abstract representation and distilled knowledge of the SDMs in order to resolve these types of problem (Fig 1 part b). As mentioned earlier, the process patterns will form a library of method chunks that can be used for assembly-based method engineering as the main application of process patterns (Fig 1 part c). They provide the reusable method chunks that help method engineer for constructing a custom SDM according to project requirements at hand.

Recently, many researchers have proposed domain specific process patterns. Ambler as stated in [2] proposed a set of activities as process patterns for developing object-oriented software applications. The process patterns called Object-Oriented Software Process (OOSP) that forms a general object–oriented SDM. Additionally, Tasharofi [10] has proposed set of process patterns extracted from a number of agile SDMs. They have identified process patterns from commonly encountered agile activities by studying seven agile SDMs. Further researches in the domain-specific patterns are existed such as *Component-Based Development* [11], *MDA-Based Development* [12], *Decision Support Systems* [13], *Aspect-Oriented* [14], and *Real Time Development* [15].

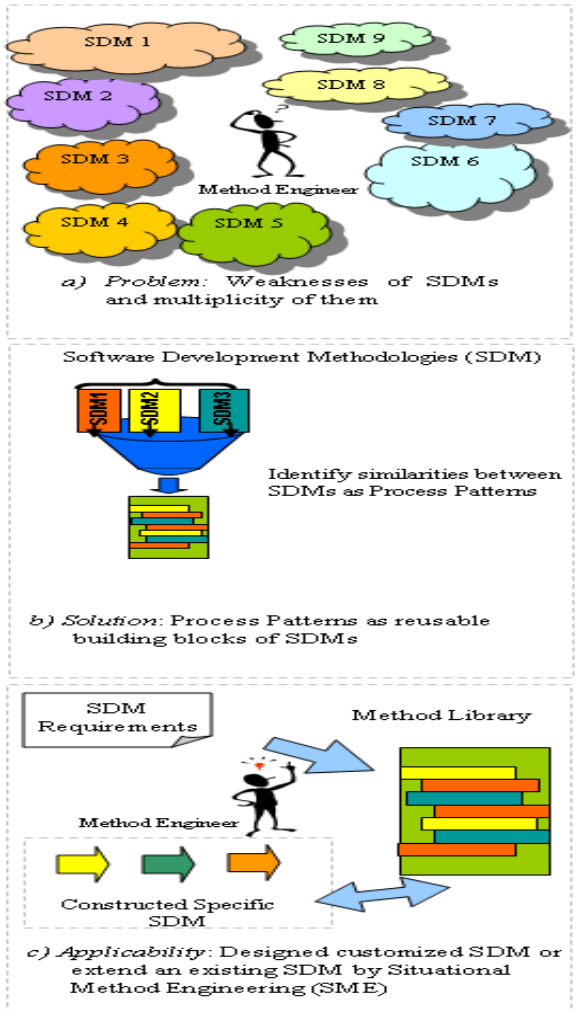

Figure 1. Positioning of process patterns in Assembly-Based Method Engineering

Although a number of process patterns have been introduced in different context, a precise and comprehensive procedure by which SDMs will be mined for recurring activities with balanced granularity and similar concept has not been previously examined. There is no procedure available for extracting process patterns from SDMs independent of its domain. Although rigorous works have been compiled for extracting design pattern instances from exiting source code to support better software maintainability and reverse engineering, they suffer from lack of adequate documents [16]. The common problem with the approaches is that they explore the source code with a number of known fixed design patterns such as Adapter, Strategy and etc. They analyze the structural representation of source code to find any meaningful structures and relations among classes as well as matching with fixed design patterns. In contrast to this, process patterns are not well-defined already in a way that it could be easily possible to find match cases as design patterns. Process patterns discovery is based on the similarity analysis that reoccurred in existing textual SDMs. Without an explicit way some process patterns may be missed or neglected. Additionally, the extracted process patterns might be highly subjective; therefore, its reusability will be highly affected. For instance, implicit extraction procedures derive various set of patterns that is highly dependent on the involved implicit experiences. Authors believe that one of the main weaknesses of related research is the lack of explicit procedure that has been used for extracting process patterns. Consequently, the main question of the paper is as follow:

*"How can method engineer extract process patterns from a number of SDMs and organize them in well-formed granularity to obtain distilled and comparable knowledge about SDMs?"*

The contribution of this problem could be a descriptive procedure to get the SDMs and construct the required patterns. The main application of this contribution is that one can use it to obtain distilled and abstract knowledge about them. For instance, one can uses the procedure for extracting process patterns from domain of *Web-Based* SDMs. A detailed description of the proposed procedure will be presented in this paper.

The rest of this paper is structured as flows: In the next section, authors present basic definition that will be used in further. Section III mentions a detailed step-by-step description of the proposed procedure. Section IV demonstrates result of applying it for extracting process patterns from a family of methodologies. Finally, section V contains conclusion and further work.

II. BASIC DEFINITIONS

In this section, the definitions of relevant terminology and its implications are introduced.

- **Map**: Many of the processes involved in SME literature are described by process models notated using the concept of map [17]. A map is described as a directed labeled graph consisting of steps representing intentions and edges representing strategies. An intention captures the notion of a task to be accomplished whereas the strategy suggests the way in which this intent can be achieved. A map always begins with the start intention and ends with the stop intention. Fig 2 shows the map of the proposed procedure for extracting process patterns.
- **Granularity of the extracted process patterns**: The paper categorizes extracted process patterns in three levels of granularity and abstraction [2]: *Phase*, *Stage* and *Task*. A task process pattern defines detailed required

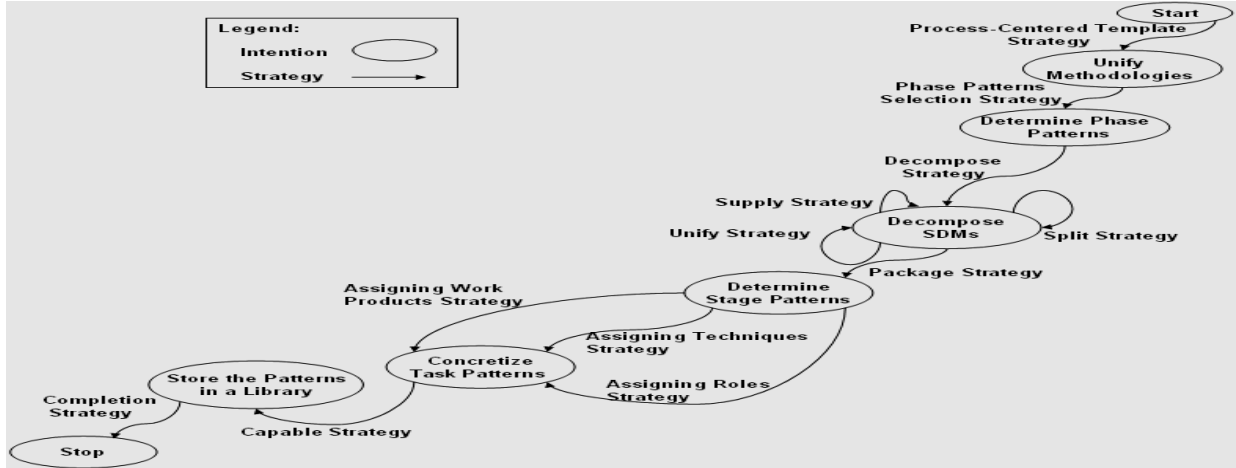

Figure2. The procedure of process patterns extraction from SDMs steps to execute a task for instance *Technical Review of Code*, and *Making Questionnaire-Interview*. Stage process pattern is contained several related tasks process patterns that need to be done to pass from a development stage to another one. Typically, they perform in iterative-incremental manner. For instance, *Develop Architecture* as a stage pattern constitutes *Design Logical Architecture*, *Design Physical Architecture* and *Evaluate Alternative Architecture* task patterns. Typically, more than one stage patterns will be placed in a phase pattern. They form a typical phase of software development lifecycle (*Construction* phase as a sample). Based on this granularity, the three intentions namely *Determine Phase Patterns*, *Determine Stage Patterns* and *Concretize Task Patterns* are applied for categorizing the patterns. Additionally, there is no constraint on the granularity levels for categorizing patterns.

• **The pattern formalism**: While the process patterns capture reusable method fragments of SDMs, the extracted patterns should be represented in a uniform and well-formed structure to facilitate their organization and maintenance. A number of formalism has been proposed to allow a better represent of process patterns such as Ambler [2], Gnatz [8], P-Sigma [18]. An overall view to these formalisms shows they have commonalities for representing a pattern. Each process pattern compromise a number of parts as shown in table I. A context is precondition that should exist while a pattern can be applied. Each pattern comes to solve concrete problems typically occurred in the given context. Process patterns can be categorized in different granularities. As mentioned above typical granularities are Phase, Stage and Task. Pattern performs by defining roles manually or by adopting case tools automatically. The input artifacts act as source to the pattern and output artifacts are result of applying the pattern.

• **Operators**: In order to mine process patterns from specific domain of SDMs, there is a need to define operators. Authors have adopted some of the generic SME operators proposed by Ralyte and Rolland [19, 20]. They are used for exploring SDMs to find similarities between activities, grouping relevant activities together and comparing them. The operators are based on the semantic similarity of activities. It should be note in the paper the operators are generic in the sense and away from real implementation. Therefore, they are accomplished manually. The generic operators are applied in the further algorithms as below:

1. **SYSNOMYM**: By this operator, similarity of two activity's names is evaluated. As illustrated example, a SDM may define an activity as *Requirements Elicitation* in which relevant data about project requirements are gathered. The *Requirements Identification* has different name with previous one, but its internal steps follows to gather customer requirements. Therefore, these activities have symmetrical mean. In this regard, they are evaluated as identical.

2. **SEMANTIC AFFINITY**: The purpose of the operator is to measure the intent closeness of two

activities. The operator groups a set of the relevant activities that have the same intent. For instance, *Making Questioner-Interview*, *Prototyping* and *Use-Case Modeling* are performed by requirement engineer to achieve a set of well-defined software requirements. While, the intent of these activities is the same, they can be considered as a group of activities relevant to requirements engineering. In contrast to that, the *Design Architecture* and *Evaluate Alternatives Architecture* in which software architecture is developed and then evaluated are divided into separate groups. In this regard, this operator can be utilized for grouping relevant activities.

TABLE I. FORMALISM FOR PATTERN REPRESENTATION [2,8,18]

| Element | | Description |
|---|---|---|
| Context | | Defines an overall situation that a problem is occurred. Artifacts are changed before and after execution of the patterns. |
| Problem | | Defines a concrete situation that may arise during system development. |
| Process Pattern | Phase Pattern | Defines detailed steps required to execute as task. |
| | Stage Pattern | Contains several task process patterns that need to be done to pass from a stage of development. Typically, it is performed in iterative-incremental manner. |
| | Task Pattern | Two or more stage patterns make a Phase patterns. |
| Roles | | Defines the person or tool that performs pattern. |
| Artifact | | Produced as the result of performing a stage or task by people or tools. |
| Related Patterns | | Relation to other process patterns such as those that use this pattern or those that can be alternative for this pattern. |
| Consequence | | List of consequences compromise by the pattern application. |

3. **MORE COMPLETE**: The operator is used to evaluate which of the two activities are more precise and complete than another. This operator is based on calculating a number of successive activities performed to reach specific goal. For instance, two SDMs may be defined a technique for designing software architecture. The former prescribes an activity by several steps to achieve the software architecture while the later only prescribes some general guidelines without any details about the required steps to design architecture merely. In this situation, the operator evaluates the former as more complete.

### III. THE PROCEDURE FOR EXTRACTING PROCESS PATTERNS

In this section, authors present the steps of the proposed procedure for extracting process patterns from SDMs. Fig. 2 represents the procedure as a map. Using the map formalism, each of its intention corresponds to one of the steps.

- *Step 1. Unify Methodologies*

Existing SDMs are rather rigid in origin, and they was not created to be modular [21]. Generally, they are represented textually in natural language so that it is too difficult to be processed by computers [7]. In addition, their modularity is limited to such an extent that they provide several models and its associated prescriptive guideline to construct different views of the software applications. Therefore, the first step is to represent SDMs in uniform structure. The "process-centered template" [9] strategy will allow the method engineer to represent uniform structure of the methodologies so that analytical comparison will become easy. The template is used for highlighting the activities prescribed in each SDM while keeping the details of the product view as secondary to the activities. The description produced using this template enables elaborate analysis of individual SDM in order to discover recurring activities that leads to identify process patterns. The structure of a SDM based on this template has been described in table II. The result of this strategy is a SDM equal to its origin with unified and comparable activities.

TABLE II. Process-Centered Template for Describing SDMs

| | | |
|---|---|---|
| Overview | | A brief introductory of SDM that distinguishes bold features, strengths, weaknesses and a visual development process that describe the SDM. |
| SDM Description | SDM's Phases | High-level sub-processes in the SDM's process consist of its activities, the order in which they are performed and a concise description of the produced work products. |
| | Details of the internal activities | Each activitiy is contained one or more steps that describe details of them. Relevant activities are placed into separate phases of the SDM. |

- *Step 2. Determine Phase Patterns*

To achieve phase process patterns, the "Phase patterns selection strategy" is used. Phase process patterns in reality represent the generic phases of Software Development Life Cycle (SDLC). In according to [22], typically a SDLC is consisting of *Initiate, Analysis and Design, Construction, Test, Deployment* and *Maintain*. It should be noted the umbrella activities have been excluded from this definition. Although details of SDMs's activities make them distinctive however at phase levels they have no considerable or innovative difference. Therefore, the strategy determines general phase process patterns as well as SDLC's phases. It should be noted in some cases, when a domain of SDMs is selected for extracting process patterns, the phases of it should be considered as phase process patterns instead of SDLC. In other cases, SDLC phases are considered as phase process patterns. While intention of the phase patterns is straightforward, a part of the selected template for representing them would be completed. For instance, table III shows a formal representation of the test phase pattern based on the proposed template. The phase process patterns work as frames for categorizing internal activities of the SDMs and will be utilized in the following stages.

TABLE III. A REPRESENTATION OF THE TEST PHASE PATTERN

| Element | Description |
|---|---|
| Context | A number of artifacts have been produced and ready to evaluate how much requirements and quality criteria has satisfied. |
| Problem | How produced artifacts can be tested? |
| Process Pattern | All stages and task patterns that included in the pattern. |
| Roles | Test engineer, test script writer, test executer. |
| Artifact | Test scripts, test results. |
| Related Patterns | This pattern corresponds to all phase patterns. |
| Consequence | To be explored. |

- *Step 3. Decompse the SDMs*

The intent of the "Decompose the SDMs" is to obtain a context for analyzing the SDMs's activities. To do this, the "Decompose strategy" helps method engineer to decompose SDMs's activities. Every activity in the underlying SDM is a candidate to be defined as a process patterns and more precisely task patterns. Having determined the phase process patterns, "Decompose strategy" decomposes SDMs and puts the internal activities to the corresponding phase process patterns. As shown in Fig 3, the different activities in the SDMs with same color have same intent and therefore fall into same phase pattern. Yellow activities in the SDMs show relevance to specific phase. For instance, it can be requirements elicitation that recurred with different names in different SDMs.

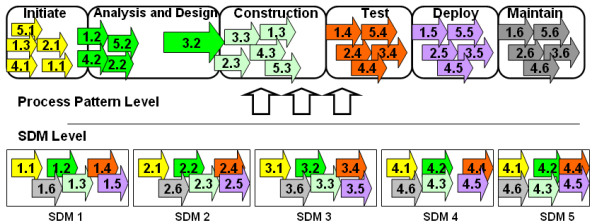
Figure 3. Decompose SDMs activities into phase process patterns

The activities may be positioned completely in a phase or mediate between two of them. According to the algorithm 1 (Fig.4), each SDM passed along the phase process patterns. The SYSNOMYM operator, as similarity analyzer, checks whether activity's phase is synonym with one or more phase process pattern. In this case, activity's phase name has different name with phase process pattern. SEMANTIC AFFINITY operator checks closeness of activity's phase and phase process pattern. Consequently, SDMs's activities are positioned in relevant phase patterns.

```
Algorithm for Decomposing Methodologies
foreach PhaseProcessPattern P_i
    foreach Methodology M_j
        foreach activity_i in M_j
            if (activity_i.phase.name SYNONYM P_i.phasename)
               &&
               (activity_i.phase.intent SEMANTIC AFFINITY P_k.intent) then
                P_i ← P_i ∪ activity_i
        endfor
    endfor
endfor
```
Figure 4. Algorithm for decomposing SDMs into phase process patterns

After decomposing the SDMs, in order to eliminate redundancy and improve cohesiveness of the activities, the map suggests three different strategies. Selection of appropriate strategy depends on situation.

▪ **Unify Strategy**: Two or more activities of SDMs might have different names but be identical semantically. By conducting **SYSNOMYM** operator, the unification of activities will be performed and only one activity will be remained. Indeed, this activity is a task pattern (figure 5.a).

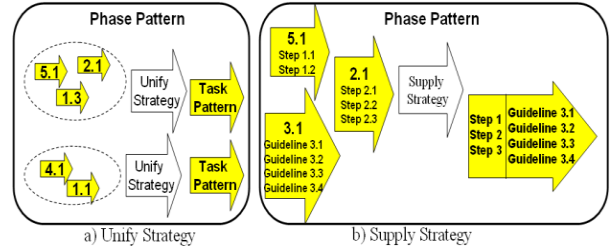
Figure 5. Algorithm of Unify Strategy

Fig 6 shows the unification algorithm.

```
Algorithm for Unify Activities
foreach PhaseProcessPattern F_i
    foreach pair of (activity_i, activity_j)
        if (activity_i.name SYNONYM activity_j.name) then
            P_i ← P_i ∪ ¬ activity_j ;
    endfor
endfor
```
Figure 6. Algorithm of Unify Strategy

▪ **Supply Strategy**: It is obvious that SDMs may prescribe different ways for performing an activity. To obtain appropriate activities in order to form a rich method fragment library, "Supply strategy" aims to capture useful parts of activities. It helps the method engineer to deal with making a complete activity. In some cases, a number of activities are synonym but one of them is more complete than the others. By adopting this strategy, these activities will be combined together to construct more complete activity with more added value. For instance, one SDM may only define one or two steps of Design Software Architecture, and the other SDM focus only on the other necessary steps of it. For instance, one focuses on primary steps of Design Software Architecture precisely, while the other SDM focus on later steps of it without mentioning the other required steps. In this situation, while these two activities can enrich each other, they should be appended. Figure 5.b shows a situation in which each of three synonym activities provides only particular steps for performing specific activities. But none of them is complete independently. Given this situation, the strategy appends them to obtain complete task pattern. Figure 7 shows the algorithm for Supply strategy.

```
Algorithm for Supply Activities
foreach PhaseProcessPattern P_i
    foreach pair of (activity_i, activity_j)
        if (activity_i.intent SYNONYM activity_j.intent)
           &&
           (activity_i.steps MORE COMPLETE activity_j.steps) then
            activity_i.step ← activity_i.step ∪ activity_j.step
            activity_i.inputartifacts ← activity_i.Inputartifacts ∪ activity_j.inputartifacts
            activity_i.outputartifacts ← activity_i.outputartifacts ∪ activity_j.outputartifacts
    endfor
endfor
```
Figure 7. Algorithm of Supply Strategy

▪ **Split Strategy**: The "Split strategy" is relevant when some activities might be too course-grained that makes

them complicated to adopt as a pattern. Therefore, these activities will be decomposed to make more appropriate activities. For instance, *Design Software Architecture* could be decomposed to *Design Logical Architecture*, *Design Technical Architecture* and *Evaluate Alternative Architecture*.

- *Step 4. Determeine Stage Patterns*

Having sieved the activities in step 3, those activities that have affinity (semantically related) to each other are grouped and make a stage pattern (Package strategy). In reality, this step clusters the relevant activities to form a group of cohesive activities as stage patterns. This is conducted by relationship analysis. This analysis will be repeated until all the activities are grouped and situated into their appropriate stage patterns as shown in Fig.8. The colored activities is based on the closeness of their intents are grouped in a cluster. As an example, the *Feasibility Analysis*, *Requirements Elicitation*, *Requirements Specification* and *Requirements Validation* activities semantically have similar intent, generally refer as *Requirement Engineering (RE)*, and consequently grouped in a separate group activities and form a stage pattern.

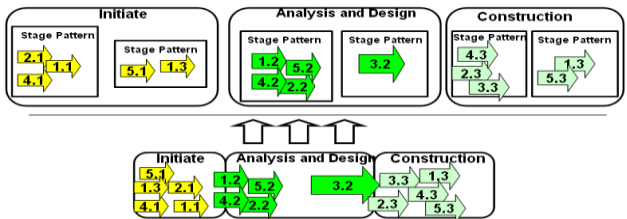

Figure 8. Grouping related activites into separate stage patterns

According to Fig.9, the SEMANTIC AFFINITY operator evaluates the intent closeness of two activities for grouping them. After making a stage pattern, a suitable name is assigned to it. SEMANTIC AFFINITY helps method engineer to provide definition for parts of the stage patterns. An example of RE stage pattern is shown in table IV.

```
Algorithm for Constitution Stage Process Patterns
while there are un staged activity do
    Create a new stage pattern S
    Pick an unstaged activity for assigning to a suitable stage pattern S ;
    S ← activity i ;
    repeat
        Pick an unstaged activity ← activity j ;
        if activity i .intent SEMANTIC AFFINITY activity j .intent then
            S ← S ∪ activity j
        until all unstaged activity has evaluated;
end
Call S as relevance stage pattern base on overall intents of the activities;
```
Figure9. Algorithm of packagin task patterns

Having conducted the relationship analysis, in some rare circumstances one activity might not be categorized in one stage pattern. Given this situation, method engineer construct an appropriate stage pattern and put this activity in it and fill the gaps (missing task patterns that have affinity to this activity) in the following steps.

TABLE IV. A REPRESENTATION OF RE STAGE PATTERN

| Element | Description |
|---|---|
| Context | A bid for new project has been offered. |
| Problem | How requirements of software can be identified and validate? |
| Process Pattern | All task patterns that included in the pattern. |
| Roles | Requirements engineer, project management. |
| Artifact | Software requirements specification, prototypes. |
| Related Patterns | To be explored. |
| Consequence | To be explored. |

- *Step 5. Concretize Task Patterns*

After elicitation of recurrent activities and grouping them into relevant stage patterns, now each task pattern should be completed based on the selected formalism. An example of task pattern has shown in table V.

It should be noted while a specific domain of SDMs has not matured enough therefore, their activities have not defined clearly and task patterns remain incomplete respectively. In this situation, it is worthwhile to enrich task patterns with ideas from prior and conventional SDMs and utilized successful practices in industry. Therefore, the appropriate technique for performing tasks should be added in an ad-hoc manner. Capitalization of experience in the other paradigm can be a starting point for this type of pattern mining. Furthermore when the SDMs provide different alternatives for doing a specific task pattern they should categorized and integrated in appropriate task adequately. For each task pattern, the two strategies "Assigning roles strategy" and "Assigning work product strategy" will complete the role and product part of the task pattern respectively.

TABLE V. A REPRESENTATION OF RI TASK PATTERN

| Element | Description |
|---|---|
| Context | A preliminary protocol agreed between stakeholders and development team for development of new system has been made. |
| Problem | How requirements of software can be gathered and identified? |
| Process Pattern | One of combination of well-know techniques such as Interviewing, JAD, Brainstorming, Concept Mapping, Sketching and Storyboarding, Use Case Modeling, Questionnaire and Checklist, Terminology Comparison can be applied. |
| Roles | Requirements engineer. |
| Artifact | Questionnaire forms, prototypes, use case models. |
| Related Patterns | Requirements specification, requirement validation. |
| Consequence | To be explored. |

- *Step 6. Store the Patterns in a Library*

While the process patterns represent best development practices for specific domain, they should be rolled out to the organization, enabling continuous process improvements. For this purpose, the "Capable strategy" suggests the method engineer import extracted process patterns into process management Computer Aided Method Engineering (CAME) tools such as EPFC [23] or RMC [24] as plug-ins to enrich existing process library. These tools provide process-engineering capabilities by supporting method engineer in documenting and deploying development process, selecting, tailoring and quick assembling process patterns for constructing specific SDM based on project needs. The SDM created with these tools can be published and deployed as web sites. The "completion Strategy" is used when the all process patterns imported to tools.

This map is actually a pattern mining procedure and after several repetitions and revisions of the patterns, a set of well-defined process patterns would be achieved. In the next section, we will show how to apply the procedure for extracting process patterns from Service-Oriented SDMs.

IV. AN APPLICATION EXAMPLE: EXTRACTING PROCESS PATTERNS FROM SERVICE-ORIENTED SDMs

In this section, authors have adopted the proposed procedure for extracting process patterns from the domain of Service-Oriented (SO) SDMs. From the methodological point of view, in SO paradigm a system is consisted as composite of services needed to address service-oriented development endeavor. As motivation for pattern extraction in a specific domain of SDMs, authors were selected twelve prominent SO-SDMs then reviewed and highlighted recurring activities. The twelve SO-SDMs that studied for extracting process patterns are: *IBM SOAD*, *IBM SOMA 2008*, *CBDI-SAE Process*, *SOUP*, *MASOM*, *SOA RQ, Papazoglou*, *RUP for SOA*, *SOAF*, *Steve Jones' Service Architectures*, *Service Lifecycle Management (SLM)*, *SOA Governance and Management Method (SGMM)* [25]. The following are the primary motivations for extracting process patterns from SO-SDMs:

- Although most of the SO-SDMs prescribe different activities with different names, they are inherently similar.
- Multiplicity and similarity of SO SDMs confounds the method engineer and software development teams to select appropriate one in order to satisfy project situation.

While each of SDM has different weaknesses and mainly focus on different issues, a generic SO-SDM as process patterns that has elicited by identifying the recurring activities can address this challenge. By following the map illustrated in the Fig 2, process patterns extracted from all of the SO-SDMs in top-down fashion have been extracted. A brief result of applying the strategies of the procedure have presented in the following subsections.

*A. Step 1. Determine Phase Patterns*

For extracting phase patterns, by reviewing the process-centered template of the SO-SDMs, it is concluded that SOMA 2008, CBDI and Papazoglou have more complete life cycle than the other SO-SDMs. Therefore, phases of these SDMs has been considered as phase patterns (Fig 10, section a). The phases act as overall frame for categorizing internal activities of all SO-SDMs.

*B. Step 2. Decompose the SDMs*

SO-SDMs' activities are passed along the phase patterns and classified into several phases. At the step, the Unify, Supply, Split strategies are applied on the activities based on their similarities, differences and analyzing the recurring activities (Fig 10, section b). For instance, Table II shows a typical activity about *Evaluating the Organization* that is repeated with different names in all of the SO-SDMs.

*C. Step 3. Determine Stage Patterns*

The stages patterns has achieved by putting relevant activities into separate groups as stage pattern (Fig 9, section c). The SEMANTIC AFFINITY operator allows making stages. The example illustrates only partially formed stage patterns. The intent of *Review Organization's IT Strategies and Objectives*, *Analyze SOA Drivers*, *Evaluate Readiness for Migration to SOA*, *Decompose Organization*, *Identify Organization's Policies and Rules* activities are close to each other and consequently grouped in a separate stage pattern called Analyze Organization stage pattern.

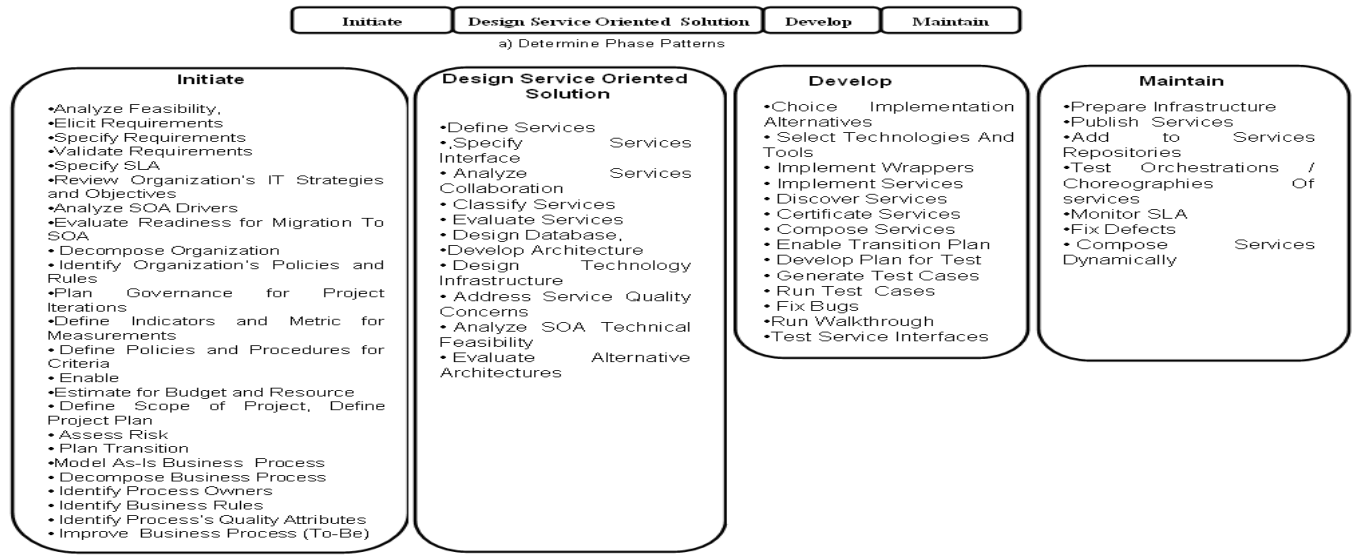
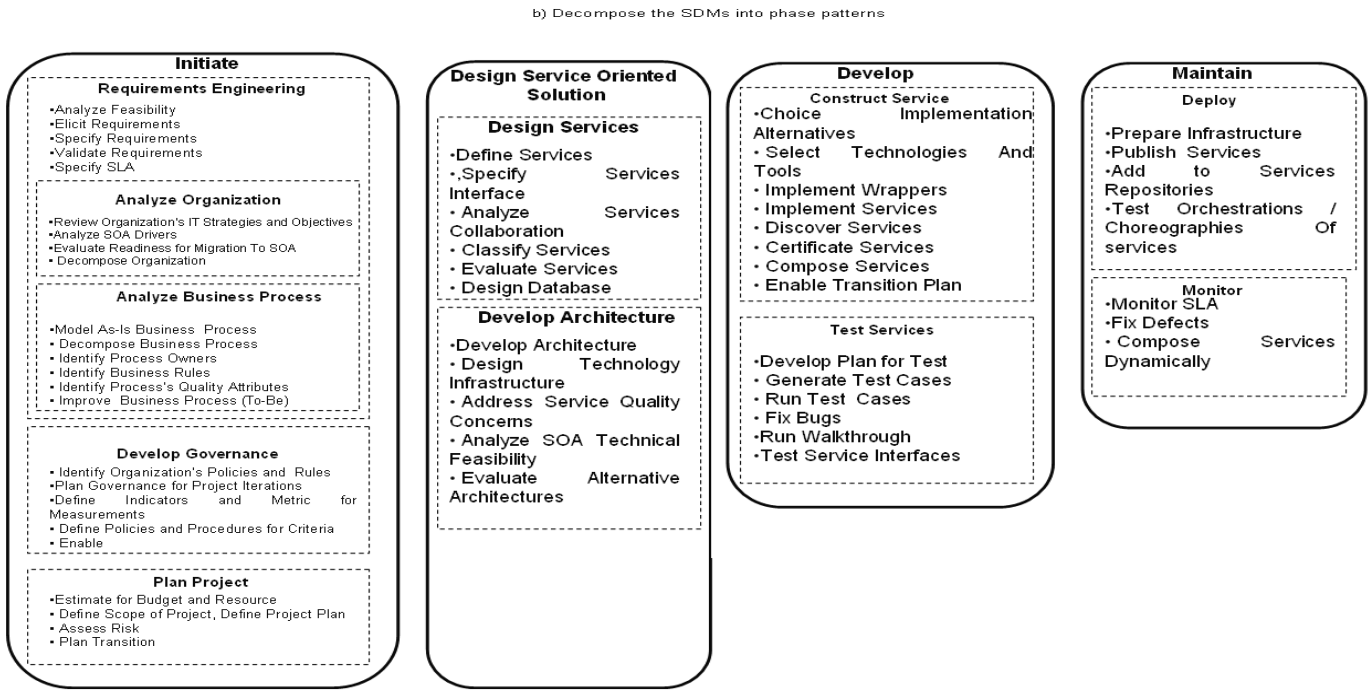
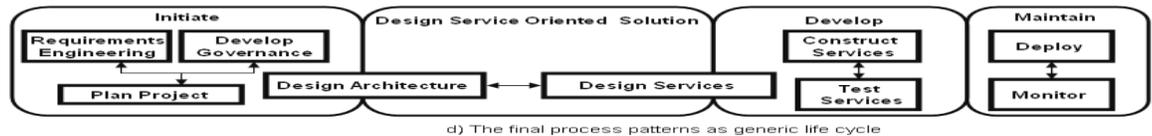

Figure 10. Steps of extracting process patterns from SO-SDMs

*D. Step 4. Concretize Task Patterns*

To have more applicable task patterns, they should be elaborated by more specific techniques. The details of task patterns can be gathered from existing SDMs or successful experiences in the service-oriented paradigm. For instance, in order to define services there are many research have proposed as Top-Down, Bottom-Up and Meet-In-the Middle. While some of the task patterns belong to more than one phases, Design Architecture and Design Services expand to Initiate and Develop phase patterns respectively. It should be noted adding patterns to process management tools are out of the scope of this research.

TABLE VI. SIMILARITIES OF SO-SDMS

| SO-SDM | Similar Activity |
|---|---|
| SOAD | Evaluate legacy systems |
| SOMA 2008 | Asset analysis |
| CBDI – SAE Process | Survey existing assets for potential services |
| SOUP | Technical infrastructure definition and analysis |
| MSOAM | Identify existing automation systems |
| RUP for SOA | Existing Asset Analysis |
| RQ | n.a |
| SOAF | Existing Application Portfolio Artifacts |
| Steve Jones | n.a |
| Papazoglou | Existing application portfolio analysis |
| SLM | Evaluate legacy systems |
| SGMM | Organization models( business entities and business processes) |

## V. CONCLUSION AND FUTURE WORK

In this paper, a novel procedure that objectively and systematically extracts process patterns from existing SDMs independent of a specific development paradigm has been proposed. To be more specific, the proposed procedure, its algorithms and operators are based on a typical data mining literature applied for analyzing similarities, and closeness of activities' intent. By adopting this procedure, quality process patterns that have comparable granularity and are independent of specific conditions of a problem domain will be achieved. The applicability of the proposed procedure has been verified in an example in which a set of specific process patterns from exiting SO-SDMs had been extracted and represented.

Although the paper provides a formal ground for processing and mining internal activities in the SDMs, however, the level of formalization and explicit definition of the concepts in SDMs are not mature enough. The existing SDMs, represented in the textbooks are narrative texts in natural language that have figures and examples for ease of understanding. However, it is difficult to process and mine them automatically [7]. In this regard, authors applied the proposed procedure and extract the patterns by human knowledge. Additionally, formalizing and implementing the proposed procedure and adding it as a plug-in in the process engineering tools to automatically populate its process repository are considered as future work.